\documentclass{elsart}

\usepackage{graphics}
\usepackage{amssymb,amsmath}
\usepackage{units}

\journal{Phys.~Lett.~B}

\newcommand{\sqrts}[1]            {$\sqrt{s}$ = \unit[#1]{GeV}}

\begin{document}

\begin{frontmatter}

\title{Forward and midrapidity like--particle ratios from $p+p$
collisions at $\sqrt{s}$=\unit[200]{GeV}}

\author[nbi]{I.~G.~Bearden},
\author[bnl]{D.~Beavis},
\author[bucharest]{C.~Besliu},
\author[newyork]{B.~Budick},
\author[nbi]{H.~B{\o}ggild},
\author[bnl]{C.~Chasman},
\author[nbi]{C.~H.~Christensen},
\author[nbi]{P.~Christiansen},
\author[krakunc]{J.~Cibor},
\author[bnl]{R.~Debbe},
\author[oslo]{E.~Enger},
\author[nbi]{J.~J.~Gaardh{\o}je},
\author[nbi]{M.~Germinario},
\author[texas]{K.~Hagel},
\author[nbi]{A.~Holm},
\author[oslo]{A.~K.~Holme},
\author[bnl]{H.~Ito},
\author[nbi]{E.~Jakobsen},
\author[bucharest]{A.~Jipa},
\author[ires]{F.~Jundt},
\author[bergen]{J.~I.~J{\o}rdre},
\author[nbi]{C.~E.~J{\o}rgensen},
\author[krakow]{R.~Karabowicz},
\author[texas]{T.~Keutgen},
\author[bnl,kansas]{E.~J.~Kim},
\author[krakow]{T.~Kozik},
\author[oslo]{T.~M.~Larsen},
\author[bnl]{J.~H.~Lee},
\author[baltimore]{Y.~K.~Lee},
\author[oslo]{G.~L{\o}vh{\o}iden},
\author[krakow]{Z.~Majka},
\author[texas]{A.~Makeev},
\author[oslo]{M.~Mikelsen},
\author[texas,kansas]{M.~J.~Murray},
\author[texas]{J.~Natowitz},
\author[nbi]{B.~S.~Nielsen},
\author[kansas]{J.~Norris},
\author[bnl]{K.~Olchanski},
\author[nbi]{D.~Ouerdane},
\author[krakow]{R.~P\l aneta},
\author[ires]{F.~Rami},
\author[bucharest]{C.~Ristea},
\author[bergen]{D.~R{\"o}hrich},
\author[oslo]{B.~H.~Samset}\footnote{Corresponding author. Email: b.h.samset@fys.uio.no},
\author[nbi]{D.~Sandberg},
\author[kansas]{S.~J.~Sanders},
\author[bnl]{R.~A.~Scheetz},
\author[nbi]{P.~Staszel},
\author[oslo]{T.~S.~Tveter},
\author[bnl]{F.~Videb{\ae}k},
\author[texas]{R.~Wada},
\author[krakow]{A.~Wieloch},
\author[bergen]{Z.~Yin} and
\author[bucharest]{I.~S.~Zgura}

(The BRAHMS Collaboration)

\address[bnl]{Brookhaven National Laboratory, Upton, New York 11973}
\address[ires]{Institut de Recherches Subatomiques and Universit{\'e} Louis
  Pasteur, Strasbourg, France}
\address[krakunc]{Institute of Nuclear Physics, Krakow, Poland}
\address[baltimore]{Johns Hopkins University, Baltimore 21218}
\address[newyork]{New York University, New York 10003}
\address[nbi]{Niels Bohr Institute, Blegdamsvej 17, University of Copenhagen, Copenhagen 2100, Denmark}
\address[krakow]{Smoluchowski Inst. of Physics, Jagiellonian University, Krak\'ow, Poland}
\address[texas]{Texas A$\&$M University, College Station, Texas, 17843}
\address[bergen]{University of Bergen, Department of Physics, Bergen, Norway}
\address[bucharest]{University of Bucharest, Romania}
\address[kansas]{University of Kansas, Lawerence, Kansas 66045}
\address[oslo]{University of Oslo, Dep. of Physics, P.b. 1048 Blindern, 0316 Oslo, Norway}

\begin{abstract}
We present a measurement of $\pi^{-}/\pi^{+}$, $K^{-}/K^{+}$ and
$\bar{p}/p$ from $p+p$ collisions at $\sqrt{s}$ = \unit[200]{GeV} over
the rapidity range $0<y<3.4$. For $p_{\rm{T}} < $ \unit[2.0]{GeV/$c$}
we see no significant transverse momentum dependence of the
ratios. All three ratios are independent of rapidity for 
$y\lesssim 1.5$ and
then steadily decline from $y \sim 1.5$ to $y \sim 3$. The
$\pi^{-}/\pi^{+}$ ratio is below unity for $y>2.0$. The $\bar{p}/p$
ratio is very similar for $p+p$ and 20\% central Au+Au collisions at
all rapidities. In the fragmentation region the three ratios seem to
be independent of beam energy when viewed from the rest frame of one
of the protons. Theoretical models based on quark--diquark breaking
mechanisms overestimate the $\bar{p}/p$ ratio up to
$y\lesssim3$. Including additional mechanisms for baryon number
transport such as baryon junctions leads to a better description of
the data.
\end{abstract}

\end{frontmatter}

\section{Introduction}
The ratios of particle production in hadronic interactions are
important indicators of the collision
dynamics~\cite{Hermann:1999;Satz:2000bn}. By comparing large and small
systems over a wide range of phase space, one can address both
reaction mechanisms in simpler systems and the properties of hot and
dense nuclear matter in large systems. A thorough understanding of
$p+p$ collisions at ultrarelativistic energies is necessary both as
input to detailed theoretical models of strong interactions, and as a
baseline for understanding the more complex nucleus--nucleus
collisions at RHIC energies. Soft particle production from
ultrarelativistic $p+p$ collisions is also sensitive to the flavor
distribution within the proton, quark hadronization and baryon number
transport. Extensive data exist near midrapidity, but less is known
about the forward rapidity region where fragmentation and isospin
effects are important.

In this Letter we present measurements of like--particle charged
hadron ratios from $p+p$ collisions at a center--of--mass energy of
$\sqrt{s}$ = \unit[200]{GeV} as a function of rapidity $y=0.5 \cdot
ln((E+p_{z})/(E-p_{z}))$ and transverse momentum $p_{\rm{T}}$, and
make a comparison with similar BRAHMS results from the 20\% most central
Au+Au collisions at the same energy. We show that the $p+p$ and Au+Au
results on pion, kaon and proton like--particle ratios are consistent
over three units of rapidity, in spite of the expected large
differences in dynamics between these systems.

In $p+p$ collisions at RHIC energies two main mechanisms for particle
production are expected. At midrapidity the Bjorken
picture~\cite{Bjorken:1982qr} predicts that particles will be formed
mainly from string fragmentation, yielding values of
antiparticle--to--particle ratios close to unity. At forward
rapidities, close to the beam rapidity ($y_{b}=5.3$ at $\sqrt{s}$ =
\unit[200]{GeV}), cross--sections are instead known to be dominated by
leading particles and projectile fragments (the fragmentation
region). This means that the conservation of charge and isospin will
become increasingly important for particle production as one approaches
$y_{b}$. The present data on $\pi^{-}/\pi^{+}$, $K^{-}/K^{+}$ and
$\bar{p}/p$ show that in $p+p$ collisions at $\sqrt{s}$ =
\unit[200]{GeV} there is a midrapidity region extending out to $y
\sim 1.5$ where the particle ratios agree with the Bjorken
picture. Above this point the ratios start to decrease,
indicating the onset of fragmentation region physics. Shifting the
ratios by the beam rapidity and comparing to lower energy data, we
find a broad rapidity range where ratios of like--particle production
are independent of the incident beam energy when viewed from the rest
frame of one of the protons (limiting
fragmentation~\cite{Benecke:1969sh}).

The traditional quark--diquark breaking picture of a $p+p$ collision
fails to reproduce baryon transport in available midrapidity data,
which has been taken as evidence for several additional mechanisms
being important at higher
energies~\cite{Kharzeev:1996sq,Capella:1996th,Liu:2003wj,Kopeliovich:1988qm}. In
this Letter we provide a comparison of different model predictions
with experimental data, which, especially away from midrapidity,
provides new constraints for calculations. We show that
the commonly used event generator PYTHIA~\cite{Sjostrand:2000wi} does
not reproduce the ratio of antiproton to proton production seen in the
data at any rapidity, while the additional hypothesis of a baryon
junction within the HIJING/B~\cite{Vance:1998vh} model yields a good
agreement with both the magnitude and rapidity dependence of the
observed $\bar{p}/p$ ratio.

\section{The analysis}
The data presented in this Letter were collected with the BRAHMS
detector system during 2001. BRAHMS consists of two movable magnetic
spectrometers and a suite of detectors designed to measure global
multiplicity and forward neutrons~\cite{Adamczyk:sq}. In addition,
eight rings of plastic scintillator tiles were used to find the
collision point and provide a minimum bias
trigger~\cite{Arsene:2004cn}. To reduce the contribution of
background events valid hits in the outer three rings were required as
part of the offline analysis.  Using a GEANT simulation with the
HIJING event generator~\cite{Wang:1991ht} as input, it was estimated that
this trigger setup saw 71$\pm$5\% of the \unit[41]{mb} $p+p$ total
inelastic cross--section. Spectrometer triggers that required hits in
several hodoscopes were used in each of the two spectrometers to
enhance the event sample of $p+p$ collisions with tracks.  For this
analysis data taken at nine angle settings with respect to the
beam were used, ranging from 90$^{o}$ to 3$^{o}$ and yielding a rapidity
coverage of $0<y<3.4$ for pions.

Identification of charged hadrons ($\pi$, $K$, and $p$) was done
primarily through time--of--flight measurements. Tracks having a
measured inverse velocity ( $\beta^{-1}$) within a $\pm 2\sigma$ band
of the theoretical value for the appropriate momentum and mass, were
selected for analysis. In the forward spectrometer where particles in
general have higher momenta, identification was also provided through
the recorded radius in a Ring Imaging Cherenkov detector, and via
momentum dependent cuts in the response of a threshold Cherenkov
detector. The details of the particle identification and analysis
methods used are similar to those described
in~\cite{Bearden:2002yb,Bearden:2001kt}, but because of the lower
particle yield our calibrated time--of--flight resolutions are worse
than for Au+Au. This mainly affects the midrapidity spectrometer,
which only has time--of--flight systems. For the present analysis a
separation of $p/K$ up to $p=$\unit[2.6]{GeV/$c$} and $K/\pi$ up to
$p=$\unit[1.2]{GeV/$c$} was achieved.

\begin{figure*}[!t]
  \resizebox{0.49\textwidth}{!}
  {\includegraphics{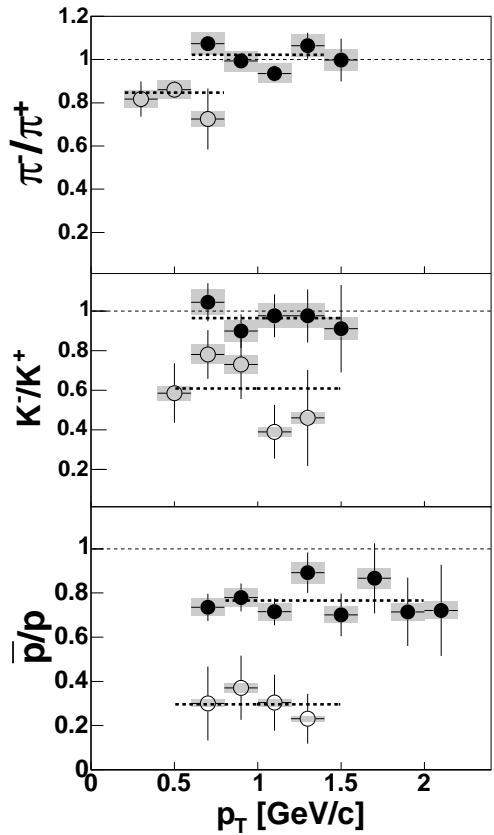}}
  \caption{\label{fig:ratiosVsPt} Particle ratios vs. $p_{\rm{T}}$ at
  $y = 0$ (solid circles) and $y \sim 3$ (open circles). The lines
  show the result of fitting a constant to the data, over the
  indicated range. The shaded area shows our estimate of the
  systematic error.}
\end{figure*}

Charged particle ratios were measured by dividing transverse momentum
spectra, normalized to the minimum bias trigger. By measuring positive
and negative particles at the same angular setting but with opposite
magnet polarities, most corrections for geometrical acceptance and
detector efficiencies cancel out. Figure~\ref{fig:ratiosVsPt} shows
the resulting like--particle ratios as a function of $p_{\rm{T}}$ at
the extreme measured rapidities of $y\sim0$ and $y\sim3$. Within our
statistical errors there is no significant dependence on
$p_{\rm{T}}$. The ratios were therefore fitted to a constant over a
$p_{\rm{T}}$ range matching the limits of our acceptance (see
Fig. 1). For most settings this range was 0.5 $<p_{\rm{T}}<$
\unit[1.5]{GeV/$c$}, varying by $< \pm$\unit[0.5]{GeV/$c$} for the
different spectrometer angles.

The ratios have been corrected for particle absorption and in--flight
decay as discussed in Ref.~\cite{Bearden:2002yb}. In addition
corrections were applied for antiproton absorption in the spectrometer
trigger slats, which removed $\sim$ 10\% of the $\bar{p}$ yield at
$p<$\unit[1]{GeV/$c$}, dropping to $\sim$ 5\% at
$p=$\unit[2]{GeV/$c$}. Primary particles were selected by by requiring
the tracks to point back to the beam line, with an achieved resolution
of $\sigma \sim 0.7$ cm. For $\pi^{-}/\pi^{+}$ and $K^{-}/K^{+}$ a
$3\sigma$ cut was used, while for $\bar{p}/p$ a $2\sigma$ cut was set
to further eliminate knock--out protons from the beampipe. Since the
spectrometers have a small solid angle the effects of feed--down from
weak decays are not large and tend to cancel in the
ratios~\cite{Bearden:2001kt}.  The $\bar{p}/p$ ratio is exceptional
since it is sensitive to the evolution with rapidity of the
$\Lambda/p$ ratio. To estimate the upper limits of this effect, a
GEANT simulation with published STAR data from $p+p$ collisions
$y=0$~\cite{Adams:2003qm;Adams:2004ec} as input has been used.  Taking
$\Lambda/p$ $\sim$ 0.5, assuming a constant behavior with rapidity and
that $\bar{\Lambda}/ \Lambda \sim \bar{p}/p \cdot K^+/K^-$ (see
e.g.~\cite{Anisovich:1972pq}), the feed-down from $\Lambda$ and
$\bar{\Lambda}$ were found to cause a net increase of $\bar{p}/p$ at
all rapidities. At midrapidity the possible contribution is $<$5\%,
and at forward rapidity $<$10\%, within our acceptance.

\section{Particle ratios vs. rapidity}
\begin{figure*}[!t]
  \resizebox{\textwidth}{!}
  {\includegraphics{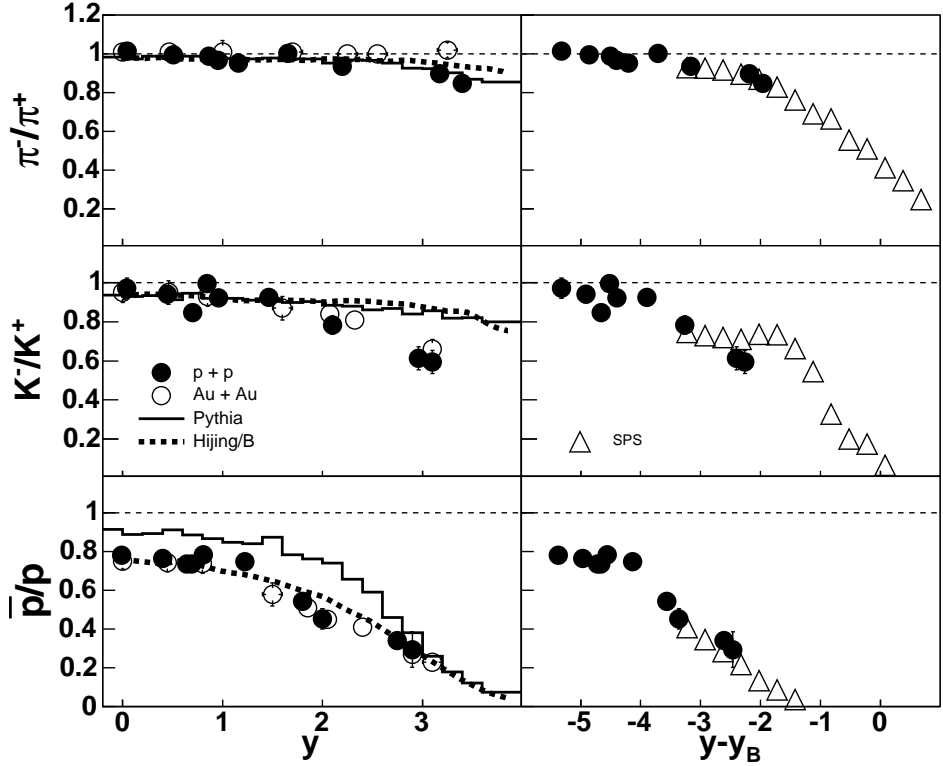}}
  \caption{\label{fig:ratiosComparison} {\bf Left:} Charged particle
    ratios from $p+p$ at $\sqrt{s}$ = \unit[200]{GeV} (solid points)
    compared with Au+Au~\cite{Bearden:2002yb} (open points), and
    predictions from PYTHIA~\cite{Sjostrand:2000wi} (solid histogram)
    and HIJING/B~\cite{Vance:1998vh} (thick dashed line). {\bf Right:}
    Ratios shifted by $y_{b}$, compared with data from NA27
    (triangles) at \sqrts{27.5}~\cite{Aguilar-Benitez:1991yy}.}
\end{figure*}

Figure~\ref{fig:ratiosComparison} shows the resulting ratios of
antiparticle--to--particle yields as a function of rapidity (left
panel). Two independent analyses were performed. By comparing these,
and by varying both the rapidity and $p_T$ intervals, and the cuts on
the particle identification and projection to the interaction point,
our point-to-point systematic errors are estimated to be $<$2\% for
pions and protons, and $<$3\% for kaons. Ratios from measurements with
different magnet polarities allow us to investigate systematic effects
from geometry and normalization. The combined residual systematic
uncertainties from these effects and from the absorption corrections are
found to be $<$5\%.

For all three ratios in Fig.~\ref{fig:ratiosComparison} there is a
clear midrapidity plateau and subsequent decrease with rapidity. The
midrapidity values of the ratios are $\pi^{-}/\pi^{+}$ =
1.02$\pm$0.01$\pm$0.07, $K^{-}/K^{+}$ = 0.97$\pm$0.05$\pm$0.07 and
$\bar{p}/p$ = 0.78$\pm$0.03$\pm$0.06, consistent within statistical
errors with values extracted from identified particle spectra reported
by STAR~\cite{Adams:2003xp}. Numbers at other rapidities are given in
Table~\ref{tab:results}. At midrapidity, proton and antiproton
production from quark--antiquark pairs can be assumed to be
identical. Proton excess, defined as
$(N_p-N_{\bar{p}})/(N_p+N_{\bar{p}})$, is therefore due to the
transport of baryon number from the initial beam. Our $\bar{p}/p$
ratio would in this interpretation imply a proton excess of 12\% at
midrapidity, carrying baryon number that has been transported from the
beam region at $y=5.3$~\cite{Kharzeev:1996sq}. We note that it has
been shown (see~\cite{Fischer:2002qp}) that one may need to correct
for isospin effects before generalizing these results from $p+p$ to
hadron--hadron collisions, due to the presence of neutrons.

\small
\begin{table*}[t]
\begin{tabular}{c|c||c|c||c|c}
\hline
\hline
Rapidity & $\pi^{-}/\pi^{+}$ & Rapidity & $K^{-}/K^{+}$ & Rapidity &
$\bar{p}/p$ \\
\hline
\hline
0.0 & 1.02$\pm$0.01 & 0.0 & 0.97$\pm$0.05 & 0.0 & 0.78$\pm$0.03 \\
0.5 & 1.00$\pm$0.01 & 0.4 & 0.94$\pm$0.04 & 0.4 & 0.76$\pm$0.03 \\
0.9 & 0.99$\pm$0.01 & 0.7 & 0.85$\pm$0.04 & 0.6 & 0.74$\pm$0.03 \\
1.0 & 0.97$\pm$0.01 & 0.8 & 1.00$\pm$0.04 & 0.7 & 0.74$\pm$0.02 \\
1.2 & 0.95$\pm$0.01 & 1.0 & 0.92$\pm$0.04 & 0.8 & 0.78$\pm$0.03 \\
1.7 & 1.00$\pm$0.01 & 1.5 & 0.93$\pm$0.03 & 1.2 & 0.75$\pm$0.02 \\
2.2 & 0.94$\pm$0.01 & 2.1 & 0.78$\pm$0.05 & 1.8 & 0.54$\pm$0.03 \\
3.2 & 0.90$\pm$0.01 & 3.0 & 0.61$\pm$0.06 & 2.0 & 0.45$\pm$0.05 \\
3.4 & 0.85$\pm$0.03 & 3.1 & 0.60$\pm$0.06 & 2.7 & 0.34$\pm$0.04 \\
    &               &    &              & 2.9 & 0.29$\pm$0.09 \\
\hline
\hline
\end{tabular}
\caption{\label{tab:results}Numerical values for charged particle
  ratios as a function of rapidity. Errors are statistical only. In
  addition a combined systematic error of 7\% for $\pi^{-}/\pi^{+}$
  and $K^{-}/K^{+}$, and 8\% for $\bar{p}/p$ is estimated.}
\end{table*}
\normalsize

At $y \lesssim 1.5$ the Au+Au ratios for the 20\% most central
collisions reported in~\cite{Bearden:2002yb} are noticeably similar to
the present results. Above $y=1.5$ the pion ratios in $p+p$ start to
drop below those for Au+Au and consequently below unity, while the
kaon and proton ratios remain consistent with the Au+Au results over
our entire acceptance range. This is surprising in view of the
different dynamics one might expect for the two systems. A heavy ion
system has multiple initial collisions as well as significant
rescattering and may reach thermal equilibrium before freezeout
occurs, while the significantly smaller $p+p$ system should not
interact much beyond the initial reactions. For all three species the
ratios start to decrease above $y=1.5$, indicating a transition from
the string breaking dominated regime at midrapidity to the
fragmentation region. The drop in the pion ratio at high rapidity can
be attributed to isospin and charge conservation in the fragmentation
region, an effect not seen for Au+Au where the high pion multiplicity
drives the system towards isospin equilibration.

The right panel of Fig.~\ref{fig:ratiosComparison} shows the present
data and data from NA27 at \sqrts{27.5}~\cite{Aguilar-Benitez:1991yy}
(open triangles) shifted by the respective beam rapidities. Overlaying
the two datasets the ratios appear to be independent of the incident
beam energy when viewed from the rest frame of one of the protons, in
the region where our rapidity coverage overlaps with that of
NA27. This is consistent with the idea of limiting fragmentation that
has also been observed for charged hadrons in nucleus--nucleus
collisions~\cite{Bearden:2001qq;Deines-Jones:1999ap;PhobosFrag}. This
hypothesis states that the excitation of the leading protons saturates
at a moderate energy, leaving more available kinetic energy for
particle production below the beam rapidity. We also note a transition
in behavior at $y-y_{b} \sim -4$, indicative of a boundary between the
midrapidity and fragmentation regions. Below this, at RHIC energies we
observe a region of constant relative particle production that was not
present at \sqrts{27.5}.

\section{Predictions from models}
To interpret these results further, predictions from theoretical
models of hadron-hadron collisions are confronted with the data.  The
curves in the left panel of Figure~\ref{fig:ratiosComparison} compare
our results to the predictions of two such calculations, PYTHIA
Ver. 6.303~\cite{Sjostrand:2000wi}\footnote{PYTHIA version 6.3 is at
the time of writing still labeled as `experimental', but we find no
difference in the results between this version and the latest in the
6.2 series.} and HIJING/B~\cite{Vance:1998vh}, using the same
$p_{\rm{T}}$ range as the present analysis.  Both models give a good
description of the pion data and for kaons at midrapidity, but do not
reproduce the magnitude of the decrease with rapidity seen for
$K^{-}/K^{+}$ as the rapidity approaches that of the fragmentation
region. Also, PYTHIA clearly overestimates the $\bar{p}/p$ ratios.
This is a well--known problem since PYTHIA employs only quark--diquark
breaking of the initial protons, while several authors have pointed
out~\cite{Kharzeev:1996sq,Capella:1996th} that to describe stopping at
midrapidity in high energy hadronic collisions one needs an additional
mechanism to transport baryon number away from the beam rapidities.

Based on $p+p$ data from the ISR it has been proposed that other
mechanisms than quark--diquark breaking, e.g. destruction of the
diquark, can transport baryon number over a large rapidity
range~\cite{Kopeliovich:1988qm}. Subsequently a description was
formulated of the baryon transport process as arising from gluonic
degrees of freedom, with an additional transport component slowly
changing with incident energy~\cite{Kharzeev:1996sq}. This can lead to
a significant net baryon content at midrapidity. Also, data from
HERA~\cite{Adloff:1998kv} show a baryon asymmetry, defined in
lepto--production as $2 \cdot (N_p-N_{\bar{p}})/(N_p+N_{\bar{p}})$,
that is significantly different from zero. This indicates that baryon
transport over 7 units of rapidity is indeed possible.  Together,
these theories and observations form the basis for implementing the
baryon junction~\cite{Kharzeev:1996sq,Rossi:1977cy}. This mechanism
allows for easy transport of baryon number toward midrapidity, while
energy balance is maintained through an increased production of
forward mesons. The baryon junction scenario, incorporated as a model
prediction in the HIJING/B event generator~\cite{Vance:1998vh}, has
successfully predicted the slow $\sqrt{s}$ dependence of the $p+p$ and
$\bar{p}+p$ cross--sections~\cite{Kharzeev:1996sq}.  In
Fig.~\ref{fig:ratiosComparison} the dashed lines showing the HIJING/B
prediction for $\bar{p}/p$ at $\sqrt{s}$ = \unit[200]{GeV}, exhibit a
much better agreement with the data than PYTHIA, both in terms of
overall magnitude and the width of the distribution.

In Ref.~\cite{Capella:sx} a baryon junction extension to a
quark--diquark breaking model of particle production is suggested. It
is shown that it is possible to describe baryon stopping in $p+p$ and
Au+Au collisions using the same parameters for the baryon junction
couplings, but with different parameter values for SPS and RHIC
energies. For RHIC, this leads to a prediction that the shapes of the
rapidity distributions for $p+p$ and Au+Au will be similar for
$|y|\lesssim2$. The similarity shown here of $\bar{p}/p$ in $p+p$ and
Au+Au up to $|y| < 3$ supports this prediction.

\begin{figure*}[!t]
  \resizebox{0.49\textwidth}{!} {
    \includegraphics{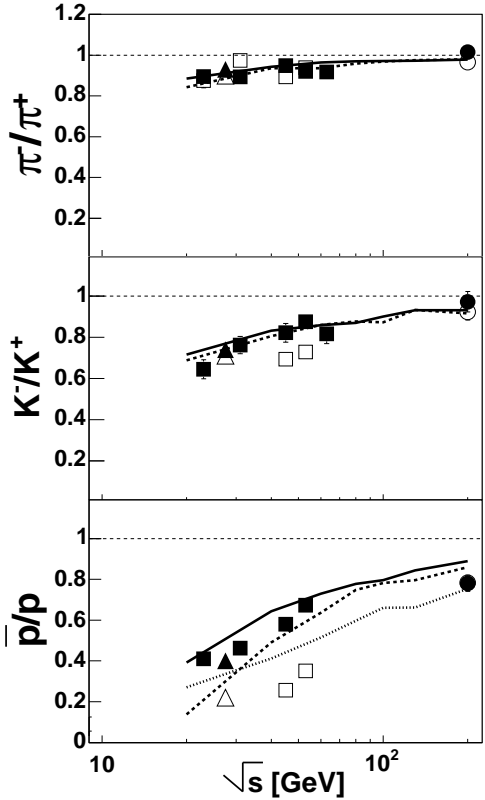}
  }
  \caption{\label{fig:energy} $\sqrt{s}$ dependence of particle ratios
    at $y=0$ (closed symbols) and $y \sim 1$ (open symbols). Circles are
    the present data, errors are statistical only.  Also shown are
    $p+p$ data from ISR (squares) and NA27
    (triangles)~\cite{Aguilar-Benitez:1991yy,Alper:1975jm}. Solid
    lines: PYTHIA prediction for $p+p$ at $y=0$. Dashed lines: same
    for $y=1$. Dotted line in bottom panel: HIJING/B prediction for
    $\bar{p}/p$ at $y=0$.}
\end{figure*}

\section{Particle ratio excitation functions}
The present data allow for an extended study of the excitation
function of the particle ratios around midrapidity. In
Figure~\ref{fig:energy} the present data at $y=0$ and $y \sim 1$ are
shown, together with fits to ISR data~\cite{Alper:1975jm} from $p+p$
collisions in the range 23 $< \sqrt{s} <$ \unit[63]{GeV}. Where
possible the fits have been made over the same $p_{\rm{T}}$ range as
our data, the notable exception being the $\bar{p}/p$ ratios at $y=1$
where the ISR data cover $2.0<p_{\rm{T}}<$\unit[4.0]{GeV/$c$}. Points
from NA27 at \sqrts{27.5} are also shown. Both at midrapidity and at
$y=1$ the ratios depend logarithmically on $\sqrt{s}$, but the slope
of this dependence is steeper at $y=1$. At lower energies there is a
significantly larger fraction of $K^{-}$ and antiprotons at $y=0$ than
at $y=1$, but this effect is much smaller at RHIC energies. This again
indicates that at RHIC there is a midrapidity source that is almost
free of net strangeness and baryon number.

The solid and dashed lines in Fig.~\ref{fig:energy} show the
prediction for the particle ratio excitation function from PYTHIA at
$y=0$ and $y=1$ respectively. At midrapidity the ratios are well
reproduced at all values of $\sqrt{s}$, except for the $\bar{p}/p$
ratio at RHIC energies, but at $y=1$ the $K^{-}/K^{+}$ and $\bar{p}/p$
do not seem well described at lower energies. The dotted line shows
the prediction for $\bar{p}/p$ from HIJING/B at $y=0$, reproducing the
result at $\sqrt{s}$ = \unit[200]{GeV} but underpredicting the results
at lower energies. For pions and kaons HIJING/B reproduces the PYTHIA
curves shown.

\begin{figure*}[!t] {
  \resizebox{0.49\textwidth}{!} {
    \includegraphics{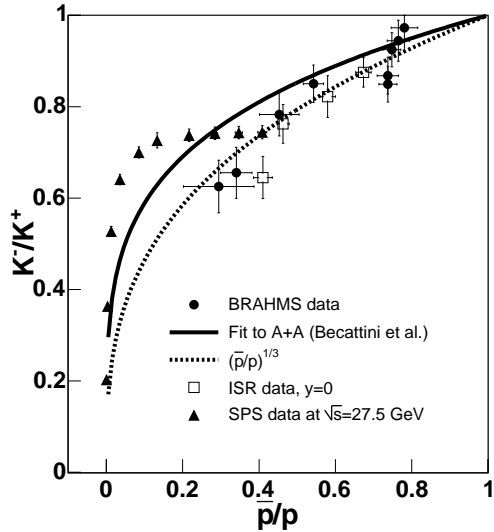}
  }
  \caption{\label{fig:thermal} Correlation between $K^{-}/K^{+}$ and
    $\bar{p}/p$ at different rapidities from the present data and data
    at lower energies. The lines show grand canonical model
    calculations for the limit of vanishing strangeness chemical
    potential $\mu_s=0$ (dashed) and for a constant temperature of
    \unit[170]{MeV} with unit strangeness
    saturation~\cite{Becattini:2000jw} (solid).}  }
\end{figure*}

\section{Ratio correlations over three units of rapidity}
For nucleus--nucleus collisions at ultrarelativistic energies it has
been observed that almost all particle production ratios can be
reproduced by a grand canonical model description of the emitting
source, i.e. with temperature $T$ and baryochemical potential $\mu_q$
as independent parameters~\cite{Braun-Munzinger:2001ip}. The strange
quark chemical potential $\mu_s$ is fixed by conservation of
strangeness~\cite{Rafelski:1991rh}. In such an approach
antiparticle--to--particle ratios are controlled by the light and
strange quark fugacities, $\mu_q/T$ and $\mu_s/T$, respectively,
predicting e.g.
\begin{equation}
K^-/K^+=e^{2\mu_s/T} \cdot e^{-2\mu_q/T} = e^{2\mu_s/T}\cdot(\bar{p}/p)^{\frac{1}{3}}
\label{GrandCon}
\end{equation}

For an ideal quark gluon plasma one can expect $\mu_s=0$, a condition
that is difficult to achieve for a hadron
gas~\cite{Letessier:1993hi;Sollfrank:1993wn}. The analysis in
Ref.~\cite{Bearden:2002yb} on data from Au+Au collisions at $\sqrt{s}$
= \unit[200]{GeV} showed that one can parametrize the kaon and proton
ratios at different rapidities as a power law: $K^{-}/K^{+} =
(\bar{p}/p)^\alpha$, with $\alpha^{Au+Au}=0.24 \pm 0.02$. Expressing
this in terms of chemical potentials gives $\mu_s \approx 0.28 \mu_q$
for Au+Au collisions.

Figure~\ref{fig:thermal} shows a similar analysis based on the present
data, where the $K^{-}/K^{+}$ ratios have been interpolated to the
same rapidities as the $\bar{p}/p$ data. A power law fit to the
present points gives an exponent of $\alpha^{p+p}=0.32 \pm 0.04$,
with $\chi^2/NDF=1.22$. Figure~\ref{fig:thermal} also shows the
corresponding results for $p+p$ collisions at \sqrts{27.5} at
rapidities $0 < y < 3.5$, and midrapidity data at ISR energies
\cite{Aguilar-Benitez:1991yy,Alper:1975jm}.  The ISR results are
consistent with the power law fit to our data, while the \sqrts{27.5}
data seem to follow a different trend.

The solid line in Fig.~\ref{fig:thermal} is the prediction of a grand
canonical calculation for a constant temperature of
\unit[170]{MeV}~\cite{Becattini:2000jw}. This curve gives a good
description of our Au+Au data, as well as lower energy heavy ion
results.  For $y < 2.0$ the $p+p$ data are also consistent with this
curve, but at more forward rapidities they fall below it. Ideally for
$p+p$ collisions one would use a microcanonical approach in order to
exactly conserve quantum numbers in each event. Such a description is
being developed e.g. by the authors of
Ref.~\cite{Liu:2003id,Becattini:2004rq}, but they also show that the
$K^{-}/K^{+}$ and $\bar{p}/p$ ratios change by $<4\%$ when going from
the canonical to the microcanonical description.

The limit of a canonical ensemble can be reached from a grand
canonical description by letting all chemical potentials approach 0.
In $e^{+}+e^{-}$ collisions such a canonical approach has been
successful in describing particle ratios~\cite{Becattini:2004rq}, but
this does not imply that such collisions constitute an ideal
quark--gluon plasma. Rather it may reflect properties of the
hadronization process. In the above grand canonical approach, a power
law exponent of $\alpha=0.33$ implies that $\mu_s=0$ (see the dashed
line in Figure~\ref{fig:thermal} and Equation~\ref{GrandCon}). The fit
made to the present data suggest that this is the case for all covered
rapidities in $p+p$ collisions at $\sqrt{s}$ = \unit[200]{GeV}.

\section{Conclusions}
In conclusion, the BRAHMS experiment has measured ratios of charged
antihadron to hadron production from $p+p$ collisions at $\sqrt{s}$ =
\unit[200]{GeV}. All ratios are independent of transverse momentum
within errors for $p_{\rm{T}} <$ \unit[2.0]{GeV/$c$}. For kaons and
protons we find an overall consistency with results from Au+Au
collisions at the same energy over three units of rapidity. The
$\pi^{-}/\pi^{+}$ ratio falls steadily below the Au+Au results for $y
= 2.0 - 3.4$, as expected from conservation of initial charge and
isospin. When viewed from the rest frame of one of the protons all
ratios seem to be independent of the projectile beam energy over a
range of at least one unit of rapidity. Models based on quark--diquark
breaking of the initial protons give a reasonable description of
$\pi^{-}/\pi^{+}$, but cannot describe our $\bar{p}/p$ ratios unless
additional mechanisms of baryon transport are invoked. Introducing a
baryon junction scheme to provide additional baryon transport to
midrapidities yields a good description of our $\bar{p}/p$ data over
our full coverage of $0<y<2.9$.

After submission we have learned about a midrapidity analysis similar
to the one presented here, made by the PHOBOS
experiment~\cite{Back:2004bk}. Their result for $\bar{p}/p$ at $y=0$
is somewhat higher than ours, but within errors the ratios reported by
PHOBOS are consistent with the ones presented in this paper.

This work was supported by the Division of Nuclear Physics of the
Office of Science of the U.S. Department of Energy under contracts
DE-AC02-98-CH10886, DE-FG03-93-ER40773, DE-FG03-96-ER40981, and
DE-FG02-99-ER41121, the Danish Natural Science Research Council, the
Research Council of Norway, the Jagiellonian University Grants, the
Korea Research Foundation Grant, and the Romanian Ministry of
Education and Research (5003/1999,6077/2000).


\end{document}